\documentstyle[11pt,leqno]{article}
\newtheorem{theorem}{Theorem}[section]

\newtheorem{proposition}[theorem]{Proposition}

\catcode`\@=11
\def\@begintheorem#1#2{\it \trivlist \item[\hskip
 \labelsep{\bf #1\ #2.}]}
\def\@opargbegintheorem#1#2#3{\it \trivlist\item[\hskip%
 \labelsep{\bf #1\ #2.\ (#3)}]}
\def\@endtheorem{\endtrivlist}

\def\@listI{\leftmargin\leftmargini \parsep 1pt plus 2.5pt
 minus 1pt\topsep 5pt plus 2pt minus 3pt%
 \itemsep 0pt plus 2.5pt minus 1pt}
\let\@listi\@listI
\@listi

\def\@sect#1#2#3#4#5#6[#7]#8{\ifnum #2>\c@secnumdepth%
 \def \@svsec {}\else \refstepcounter {#1}\edef \@svsec%
 {\csname the#1\endcsname. \hskip .1em }\fi \@tempskipa%
 #5\relax \ifdim \@tempskipa >\z@ \begingroup #6\relax%
 \@hangfrom {\hskip #3\relax \@svsec }{\interlinepenalty%
 \@M #8.\par }\endgroup \csname #1mark\endcsname {#7}%
 \addcontentsline {toc}{#1}{\ifnum #2>\c@secnumdepth%
 \else \protect \numberline {\csname the#1\endcsname. }%
 \fi #7}\else \def \@svsechd {#6\hskip #3\@svsec #8.%
 \csname #1mark\endcsname {#7}\addcontentsline {toc}{#1}%
 {\ifnum #2>\c@secnumdepth \else \protect \numberline%
 {\csname the#1\endcsname. }\fi #7}}\fi \@xsect {#5}}

\def\section{\@startsection {section}{1}{\z@ }%
 {-3.5ex plus -1ex minus -.2ex}{2.3ex plus .2ex}{\bf }}

\def\thebibliography#1{%
 \section *{References.\@mkboth {REFERENCES}{REFERENCES}}%
 \list {[\arabic {enumi}]}{\settowidth \labelwidth {[#1]}%
 \leftmargin \labelwidth \advance \leftmargin \labelsep %
 \usecounter {enumi}} \def \newblock %
 {\hskip .11em plus .33em minus -.07em} \sloppy \clubpenalty 4000%
 \widowpenalty 4000 \sfcode`\.=1000\relax}

\catcode`\@=13

\def\qed{\hspace*{\fill}
\mbox{\hphantom{mm}\rule{0.25cm}{0.25cm}}\\}

\def\boldk{\bf k}
\def\tp#1#2{{#1}^{\otimes{#2}}}
\def\ident{1\!\!1}
\def\b{\bullet}
\def\bod{\bigodot}
\def\bv{{\cal B}(V)}
\def\db{d_{\cal B}}
\def\calb{{\cal B}}
\def\domega{{d_{C}}}
\def\dcobar{\domega}
\def\der#1{\mbox{\rm Der}^{#1}_V(\bod(V,\bv))}
\def\br#1{{\mbox{\rm Br}_{#1}}}
\def\T#1{{\cal T}_{#1}}
\def\ck#1#2{C_{#1}(K_{#2})}
\def\int#1{\{0,\ldots,#1\}}
\def\cdots{\!\cdot\!\cdot\!\cdot\!}
\def\bolde{{\bf e}}
\def\bolds{{\bf s}}
\def\boldt{{\bf t}}
\def\boldu{{\bf u}}
\def\hom#1#2#3#4{\mbox{\rm Hom}_{#1}^{#4}(#2,#3)}
\def\HOM#1#2#3{\mbox{\bf Hom}^{#3}(#1,#2)}

\def\dhoch{d_{\mbox{\scriptsize\rm Hoch}}}
\def\dqcohoch{d_{\mbox{\scriptsize\rm q-coH}}}

\def\od{\odot}
\def\susp{\uparrow\hskip-1mm}
\def\E#1#2{{E^{#1,#2}(A)}}
\def\mu{\hskip.2mm\cdot\hskip.8mm}

\begin{document}
\bibliographystyle{plain}
\baselineskip15pt

\hphantom{m}
\vskip1cm
\begin{center}
\Large\bf
Drinfel'd algebra deformations and the associahedra%
\footnote[1]{First author partially supported by the
E.~Schr\"odinger International Institute for Mathematical Physics,
Vienna, and by the National Research Counsel, USA}
\end{center}

\begin{center}
\large Martin Markl and Steve Shnider
\end{center}

\vskip7mm
\begin{center}
\begin{minipage}{16.3cm}
\baselineskip13pt
{\bf Introduction.}
The aim of this work is to construct a cohomology theory
controlling the deformations of a general Drinfel'd algebra $A$ and
thus
finish the program which began in~\cite{markl-stasheff:preprint},
\cite{shnider-sternberg:preprint}.
The picture presented here has two sides -- the combinatorial one
related with the fact of the existence of a differential graded
Lie algebra
structure on the simplicial cochain complex of the associahedra, and
the algebraic one related with the algebra of derivations on the bar
construction. The relationship between these two parts is described
in Theorem~\ref{existence_of_homomorphism}. A full exposition of
the results announced here will be given in~\cite{MS:final}.
\end{minipage}
\end{center}

\baselineskip17pt
\section{Algebraic preliminaries}

Recall that a Drinfel'd algebra (or a quasi-bialgebra in the
original terminology of~\cite{D}) is an object
$A=(V,\mu,\Delta,\Phi)$, where
$(V,\mu,\Delta)$ is an associative, not necessarily coassociative,
unitary $\boldk$-bialgebra, $\Phi$ is an invertible element of
$\tp V3$,
and the usual coassociativity property is replaced by
\[
(\ident\otimes\Delta)\Delta \cdot \Phi
= \Phi\cdot(\Delta\otimes\ident)\Delta,
\]
where we use the dot $\mu$ to indicate both the (associative)
multiplication on $V$ and the induced multiplication on $V^{\otimes
3}$.
Moreover, the validity of the following
``pentagon identity'' is required:
\[
(\ident^2\otimes\Delta)(\Phi)\cdot(\Delta\otimes\ident^2)
(\Phi) =
(1\otimes\Phi)\cdot(\ident
\otimes\Delta\otimes\ident)(\Phi)\cdot(\Phi\otimes1).
\]

For a $(V,\mu)$-module $N$, define the following generalization of
the $M$-construction of \cite[par.~3]{markl-stasheff:preprint}.
Let $F^*=\bigoplus_{n\geq 0}F^n$ be
the free unitary nonassociative $\boldk$-algebra generated by
$N$, graded by length of words. The algebra $F^*$ admits a
natural left action,
$(a,f)\mapsto a\b f$, of $(V,\mu)$ given by the rules:
\begin{itemize}
\item[(i)]
on $F^1=N$, the action is given by the action of $V$ on $N$
and
\item[(ii)]
$a\b (f\star g)= \sum(\Delta'(a)\b f)\star(\Delta''(a)\b g)$,
\end{itemize}
where $\star$ stands for the multiplication in $F^*$ and we use the
notation $\Delta(a)= \sum \Delta'(a)\otimes \Delta''(a)$. The right
action $(f,b)\mapsto f\b b$ is defined by the similar rules. These
actions define on $F^*$ the structure of a $(V,\mu)$-bimodule.

Let $\sim$ be the relation, $\star$-multiplicatively generated on
$F^*$ by the expressions of the form
\begin{equation}
\label{relation}
\left((\Phi_1\b x)\star\left((\Phi_2\b
y)\rule{0mm}{4mm}\star(\Phi_3\b
z)\right)\right)
\sim\left(\left((x\b\Phi_1)\star\rule{0mm}{4mm}
(y\b\Phi_2)\right)\star
(z\b\Phi_3)\right),
\end{equation}
where $\Phi = \sum\Phi_1\otimes \Phi_2\otimes\Phi_3$ and $x,y,z\in
F^*$.
Put $\bigodot(N):= F/\sim$.
Just as in~\cite[Proposition~3.2]{markl-stasheff:preprint}
one proves that the $\b$-action induces on $\bigodot(N)$ the structure of
a $(V,\mu)$-bimodule (denoted again by~$\b$) and that $\star$ induces
on $\bigodot(N)$ the nonassociative multiplication denoted by
$\odot$. The operations are related by
\begin{equation}
\label{opulka}
a\b (f\od g)= (\Delta'(a)\b f)\od(\Delta''(a)\b g)
\mbox{ and }
(f\od g)\b b = (f\b \Delta'(b))\od (g \b \Delta''(b)),
\end{equation}
for $a,b\in V$ and $f,g\in \bigodot(N)$. Since the defining
relations~(\ref{relation}) are homogeneous with respect to the
length,
the grading of $F^*$ induces on $\bod(N)$ the grading $\bod^*(N)=
\bigoplus_{i\geq 0}\bod^i(N)$. If $N$ itself is a graded vector
space, we have also the obvious second grading, $\bod(N) =
\bigoplus_{j}\bod(N)^j$, which coincides with the first grading if
$N$ is concentrated in degree~1.

Let $\mbox{\rm Der}^n_V(\bod N)$ denote the set of degree $n$
$(V,\mu)$-linear derivations of the (nonassociative) graded algebra
$\bod(N)^*$. It is immediate to see that there is an one-to-one
correspondence between the elements $\theta \in\mbox{\rm
Der}^n_V(\bod N)$
and $(V,\mu)$-linear homogeneous degree~$n$ maps $f:N^*\to
\bod(N)^*$.

If $N=X\oplus Y$, then $\bod(X\oplus Y)$ is naturally
bigraded, $\bod^{*,*}(X\oplus Y)= \bigoplus_{i,j\geq
0}\bod^{i,j}(X\oplus Y)$, this bigrading being defined by saying that
a monomial $w$ belongs to $\bod^{i,j}(X\oplus Y)$ if there are exactly
$i$
(resp.~$j$) occurrences of the elements of $X$ (resp.~$Y$) in $w$.

Let $(\bv,\db)$ be the two-sided bar construction on the algebra
$(V,\mu)$,
i.e.~the graded differential space $\bv = \bigoplus_{n\geq
1}\calb_n(V)$, where $\calb_n(V)$ is the free $(V,\mu)$-bimodule on
$V^{\otimes n}$, i.e.~the vector space $V^{\otimes(n+2)}$ with the
action given by
\[
u\cdot (a_0\otimes\cdots\otimes a_{n+1}):= (u\cdot
a_0\otimes\cdots\otimes a_{n+1})\
\mbox{ and }\
(a_0\otimes\cdots\otimes a_{n+1})\cdot w := (a_0\otimes\cdots\otimes
a_{n+1}\cdot w)
\]
for $u,v,a_0,\ldots,a_{n+1}\in V$. If we use the more compact
notation (though a nonstandard one), writing $(a_0|\cdots|a_{n+1})$
instead of $a_0\otimes\cdots\otimes a_{n+1}$, the differential $\db
:
\calb_n(V)
\to \calb_{n-1}(V)$ is defined as
\[
\db(a_0|\cdots|a_{n+1}):=
\sum_{0\leq i\leq n}(-1)^{i}(a_0|\cdots|a_i\cdot a_{i+1}|
\cdots|a_{n+1}).
\]
Notice that the differential $\db$ is a $(V,\mu)$-linear map.  We
have two more $(V,\mu)$-linear maps, namely the `coactions' $\lambda
:\bv \to V\odot \bv \mbox{ and } \rho : \bv \to \bv\odot V$ given by
\begin{eqnarray*}
\lambda(a_0|\cdots|a_{n+1})&:=&
\sum\Delta'(a_0)\cdots\Delta'(a_{n+1})\od
(\Delta''(a_0)|\cdots|\Delta''(a_{n+1})),\mbox{ and}
\\
\rho(a_0|\cdots|a_{n+1})&:=&\sum
(\Delta'(a_0)|\cdots|\Delta'(a_{n+1}))\od\Delta''(a_0)\cdots
\Delta''(a_{n+1}).
\end{eqnarray*}

The following notation will be useful in the
sequel. Let $\susp V$ denotes the vector space $V$ considered as a
graded vector space concentrated in degree~1 and let $\tilde\bv$ be
the bar construction $\bv$ with the ``opposite'' grading,
$(\tilde\bv)^i ={\calb}_{-i}(V) $. Then put $\bod(V,\bv):=\bod(\susp
V\oplus\tilde\bv)$.
The maps above determine the following two important elements of
$\der1$: the derivation $D_{-1}$
given by $D_{-1}|_{\bv}=\db$ and $D_{-1}|_V=0$, and the derivation
$D_0$ given by $D_0|_{{\cal B}_n(V)}= \lambda+
(-1)^{n+1}\rho$, $1\leq n$, and $D_0|_V=\Delta$.

\begin{theorem}
\label{existence_of_h-structure}
There are derivations $D_k\in \der1$, $k\geq 1$, such that
$D_k(\bod^{i,j}(V,\calb_n(V)))\subset
\bod^{i+k+1,j}(V,\calb_{n+k}(V))$
for $i,j\geq 0$, $k,n\geq 1$, and that $D:=
D_{-1}+D_0+\sum_{k\geq 1}D_k$ has square zero, $D^2=0$.
\end{theorem}
The construction of $D_k$ is described in the next section.

\section{On the Stasheff associahedra and the higher derivations $D_k$}

Let $K_n$ be, for $n\geq 2$, the Stasheff
associahedron~\cite{stasheff:TAMS63}.
The vertices of $K_n$ are indexed by the set $\br n$ of full
bracketings of $n$ indeterminates. For $a,b\geq 2$ and
$1\leq t\leq b$, there is a map $(-,-)_t: \br a \times \br b
\to \br{a+b-1}$, given by inserting the first argument to the t-th
position in the second argument. This map defines the inclusions
$\iota_t :K_a\times K_b \hookrightarrow \partial K_{a+b-1}$ and it is
known that
\begin{equation}
\label{boundary}
\partial K_n = \textstyle\bigcup_J \iota_t (K_a \times K_b),
\end{equation}
where $J=\{(a,b,t);\ a+b=n+1, a,b\geq 2 \mbox{ and }1\leq t\leq
b\}$.

Let $\prec$ be the partial order on $\br n$ defined by saying that
$u\prec v$ if and only if $v$ is obtained from $u$ by the
substitution $(w_1,w_2)w_3\mapsto w_1(w_2,w_3)$ with some
$w_i\in \br{n_i}$, $i=1,2,3$, $n_1+n_2+n_3\leq n$. Let $\xi_n$ be
the (unique) minimal element of $\br n$.
\begin{proposition}
\label{observation}
\[
K_n = \mbox{\rm Cone}
(\textstyle\bigcup_{J'}\iota_t (K_a \times K_b);\xi_n),
\]
where $J'=\{(a,b,t) \in J;\ t\geq 2\}$.
\end{proposition}

The cone structure above induces on $K_n$ inductively the
triangulation $\T n$ as follows:
\begin{enumerate}
\item[(i)]
$\T2$ is the only possible triangulation of $K_2=\{\b\b\}$ and
\item[(ii)]
for $n\geq 2$, $\T n$ is induced by the cone structure
of Proposition~\ref{observation}
from the triangulations $\T a\times \T b$ of  $K_a\times K_b$,
$a+b=n+1$, $a,b\geq 2$,  which are already defined, by the induction
assumption.
\end{enumerate}
It is possible to show that $\T n$ has the property that $\prec$
induces a total order on the vertices of any simplex
of $\T n$. We will consider $\T n$ as an oriented triangulation with
the orientation induced by $\prec$.

Let $\ck in$ denote the set of $i$-dimensional oriented
simplicial chains of
$\T n$ with coefficients in $\boldk$ and let $d_S
:\ck in\to \ck {i-1}n$ be the simplicial differential.
For $\bolds\in \ck pa$ and $\boldt \in \ck qb$, $p,q \geq 0$,
$a,b\geq 2$, set
\[
\bolds\diamond\boldt:=
\sum_{1\leq t\leq b}(-1)^{(a+1)(t+q+1)} ({\iota_t})_*
(\bolds\times\boldt)\in \ck {p+q}{a+b-1}.
\]
Let $C(K)^n:=\bigoplus_{n=i-p-1}\ck pi$ and extend $\diamond$ to a
bilinear
map (denoted by the same symbol)
\mbox{$\diamond :C(K)^*\otimes C(K)^*\to C(K)^*$.}

\begin{proposition}
\label{pre-Lie}
$(C(K)^*,\diamond,d_S)$ is a graded differential
pre-Lie algebra (an analog of
the corresponding nondifferential notion of~\cite{G}), i.e.
\begin{eqnarray*}
\bolds\diamond(\boldt\diamond
\boldu)-(\bolds\diamond\boldt)\diamond\boldu &=&
(-1)^{|\bolds|\cdot|\boldt|}(\boldt\diamond(\bolds\diamond
\boldu)-(\boldt\diamond\bolds)\diamond\boldu )
\mbox{ and }
\\
d_S(\bolds \diamond \boldt) &=& d_S \bolds \diamond \boldt+(-1)^{|\bolds|}\cdot
\bolds \diamond d_S \boldt
\end{eqnarray*}
for any $\bolds,\boldt,\boldu \in C(K)^*$.
\end{proposition}

Let $[\bolds,\boldt]:= \bolds \diamond\boldt -
(-1)^{|\bolds|\cdot|\boldt|}\boldt\diamond\bolds$.
An easy computation based on Proposition~\ref{pre-Lie} shows that
$(C(K)^*,[-,-],d_S)$ is a differential graded Lie algebra.
Observe that $\der*$ is also a differential graded
Lie algebra, the bracket being given by the (graded) commutator of
derivations and the differential defined by
$\nabla(\theta):= [D_{-1}, \theta]$.

Let $\bolde_n\in \ck n{n+2}\subset C(K)^1$, $n\geq 0$, be a sequence
of `fundamental classes' chosen in such a way that
$\bolde_0=1\cdot (\b,\b)$ and that
\begin{equation}
\label{fundamental}
\mbox{$d_S \bolde_n +\sum_{i+j=n-1}(\bolde_i\diamond \bolde_j) = 0$, or,
equivalently, $d_S\bolde_n+\frac12 \sum_{i+j=n-1}[\bolde_i,\bolde_j]$}.
\end{equation}
It follows
from~(\ref{boundary}) and from the asphericity of $K_n$ that such a
sequence exists.

\begin{theorem}
\label{existence_of_homomorphism}
There exists a homomorphism
$m\! :(C(K)^*,[-,-],d_S)\to (\der*,[-,-],\nabla)$
of differential graded Lie algebras such that
$m(\bolde_0)=D_0$. Moreover, the homomorphism $m$ has the property
that
\[
\mbox{
$m(\ck pn)(\bod^{i,j}
(V,\calb_q(V)))\subset\bod^{i+n-1,j}(V,\calb_{p+q}(V)),\
i,j,p\geq 0,\ n\geq 2,\ q\geq 1.$}
\]
\end{theorem}

\begin{theorem}
The derivations $D_k:= m(\bolde_k)$, $ k\geq 1$, satisfy the
assumptions of Theorem~\ref{existence_of_h-structure}.
\end{theorem}

\noindent
{\bf Sketch of proof.} Looking at the homogeneous components of $D^2$
we see that $D^2=0$ is equivalent to $[D_{-1},D_k]+
\frac12\sum_{i+j=k-1}
[D_i, D_j] = 0$ which is, by the definition of the
derivations $D_k$ and Theorem~\ref{existence_of_homomorphism}, the
same as $m(d_S\bolde_{k}+\frac12\sum_{i+j=k-1}
[\bolde_{i},\bolde_{j}]) = 0$, which follows
from~(\ref{fundamental}).
\qed

\section{Deformation theory for Drinfel'd algebras}

As we have already observed, both $\bod(V,\bv))^*$ and $\bod^*(V)$
($=\bod(\susp V)^*$) are graded $(V,\mu)$-modules (via the
$\bullet$-action). The $\odot$-multiplication yields on both
$\bod(V,\bv))^*$ and $\bod^*(V)$ the structures of graded
$\bod^*(V)$-bimodules and both structures are related
by~(\ref{opulka}).
Denote by $\HOM{\bod(V,\bv)^*}{\bod^*(V)}n$ the set of all degree $n$
homogeneous maps $f:\bod(V,\bv)^*\to\bod^*(V)$
which are both $\bod^*(V)$ and $(V,\mu)$-linear. For such a map put
$d(f):=
f\circ D+(-1)^n \dcobar \circ f$, where $D$ is as in
Theorem~\ref{existence_of_h-structure} and $\dcobar$ is the degree
one
derivation on $\bod^*(V)$ defined by $\dcobar|_V:=\Delta$.

\begin{theorem}
\label{main_theorem}
The deformation theory of a Drinfel'd algebra $A=(V,\mu,\Delta,\Phi)$
is
controlled by the cohomology of the complex
\[
\mbox{$
\left(\HOM{\bod(V,\bv)^*}{\bod^*(V)}*,d
\right).$}
\]
\end{theorem}

Using the fact that $\bod(V,\bv)^*$ is, in a sense, the free
$\bod^*(V)$-bimodule on the $(V,\mu)$-bimodule $\tilde\bv^*$
we may write
\[
\mbox{$\HOM{\bod(V,\bv)^*}{\bod^*(V)}n \cong
\hom V{\tilde\bv^*}{\bod^*(V)}n,$}
\]
where $\hom V{\tilde\bv^*}{\bod^*(V)}n$ denotes the set of
$(V,\mu)$-linear maps $f:\tilde\bv \to \bod^*(V)$.
Using the fact that $\tilde\bv^*$ is the free $(V,\mu)$-module on
$\bigotimes^{-*}(V)$ we get the identification
\[
\mbox{$\hom V{\tilde\bv^*}{\bod^*(V)}n\cong
\hom{\boldk}{\bigotimes^{-*}(V)}{\bod^*(V)}n,$}
\]
where $\hom{\boldk}{\bigotimes^{-*}(V)}{\bod^*(V)}n$ denotes
the set of degree $n$ $\boldk$-linear homogeneous maps $g:
\bigotimes^{-*}(V)\to \bod^*(V)$. Finally, using the canonical
isomorphism $J:\bod^*(V)\cong \bigotimes^*(V)$ of graded vector
spaces constructed in~\cite[Proposition~3.3]{markl-stasheff:preprint}
we get
\[
\mbox{$\hom{\boldk}{\bigotimes^{-*}(V)}%
{\bod^*(V)}n\cong\hom{\boldk}%
{\bigotimes^{-*}(V)}{\bigotimes^*(V)}n,$}
\]
where we denote the corresponding differential again by $d$.
This is the underlying vector space of
the Gerstenhaber-Schack complex \cite{GS}
which controls the deformation of coassociative bialgebras.
In fact, our complex can be considered as a kind of deformation of the
Gerstenhaber-Schack complex as follows.
In order to understand better the structure of the differential graded space
$(\hom{\boldk}{\bigotimes^{-*}(V)}{\bigotimes^*(V)}n,d)$, let us
simplify the notation by setting
\[
\mbox{$C^n(A):=
\hom{\boldk}{\bigotimes^{-*}(V)}{\bigotimes^*(V)}n\mbox{ and }\E ij:=
\hom{\boldk}{\bigotimes^{i}(V)}{\bigotimes^j(V)}{}.$}
\]
Then it is
immediate to see that $C^n(A) = \bigoplus_{i+j=n}\E ij$ and that
the differential $d$ decomposes as $d=\dhoch +\dqcohoch
+\sum_{k=1}^{\infty}M^k$, where
\begin{eqnarray}
f\mapsto f\circ D_{-1}\quad\mbox{ induces}\quad\dhoch:\E ij
\rightarrow\E {i+1}j,\nonumber\\
f\mapsto f\circ D_0 +(-1)^n d_C\circ f\quad\mbox{ induces}\quad
 \dqcohoch:\E ij\rightarrow \E i{j+1}\label{zebrulka}\\
\mbox{ and } \quad f\mapsto f\circ D_k\quad \mbox{ induces}
\quad M^k:\E ij\rightarrow\E{i-k}{j+k+1}.\nonumber
\end{eqnarray}
Here $\dhoch$ is the Hochschild differential of the associative
algebra $(V,\mu)$, $\dqcohoch$ is the analog of the coHochschild
differential for the (noncoassociative) coalgebra $(V,\Delta)$,
and the maps $M^k$ correspond to the
derivations $D_k$ of Theorem~\ref{existence_of_h-structure}, for
$k\geq1$.
In general, when $\Phi\neq 1$, we do
not have $\dqcohoch^2=0$. However when $\Phi=1$, $\dqcohoch^2=0$ and
$\dhoch +\dqcohoch$ is the differential defined in \cite{GS}. Moreover,
in this case the operators $D_k$ vanish when restricted to normalized
cochains, so the cohomology of our complex is the same as that of
the Gerstenhaber-Schack complex.

  Bigraded differential complexes where the differential
decomposes in as in~(\ref{zebrulka}) are sometimes called
hypercompexes (although the first author prefers to call them
monsters).  Theorem~\ref{main_theorem} can be then reformulated as
\vskip2mm
\noindent
{\bf Theorem~\ref{main_theorem}.'} {\em
The deformation theory of a Drinfel'd algebra $A=(V,\mu,\Delta,\Phi)$
is
controlled by the cohomology of the hypercomplex
\[
\mbox{$(C^*(A) = \bigoplus_{i+j=*}\E ij,
d=\dhoch +\dqcohoch
+\sum_{k=1}^{\infty}M^k)$}
\]
}

\hphantom{h}
\vskip-2cm
\hphantom{h}

\baselineskip13pt
\vskip3mm
\catcode`\@=11
\noindent
M.~M.: Mathematical Institute of the Academy, \v Zitn\'a 25, 115 67
Praha 1, Czech Republic,\hfill\break\noindent
\hphantom{M.~M.:} email: {\bf markl@earn.cvut.cz}\hfill\break\noindent
\hphantom{M.~M.:} Current address (until March 1994): Math-UNC,
Chapel Hill, NC 27599-3250, USA

\noindent
S.~S.: Department of Mathematics, University of Bar Ilan, Israel,
\hfill\break\noindent
\hphantom{S.~S.:} email: {\bf shnider@bimacs.cs.biu.ac.il}


\frenchspacing
\begin{thebibliography}{10}
\baselineskip14pt

\bibitem{D}
V.G.~Drinfel'd.
\newblock Kvazichopfovy algebry.
\newblock {\em Algebra i Analiz}, {\bf 1,6}(1989), 114--148.

\bibitem{G}
G.~Gerstenhaber.
\newblock The cohomology structure of an associative rings.
\newblock {\em Annals of Mathematics}, {\bf 78,2}(1963), 268--288.

\bibitem{MS:final}
M.~Markl and S.~Shnider.
\newblock Drinfel'd algebra deformations, homotopy comodules and the
associahedra.
\newblock To appear.

\bibitem{markl-stasheff:preprint}
M.~Markl and J.D. Stasheff.
\newblock Deformation theory via deviations.
\newblock {\em Journal of Algebra}, to appear.

\bibitem{shnider-sternberg:preprint}
S.~Shnider and S.~Sternberg.
\newblock The cobar resolution and a restricted deformation theory
for
  {Drinfeld} algebras.
\newblock {\em Journal of Algebra}, to appear.

\bibitem{stasheff:TAMS63}
J.D. Stasheff.
\newblock Homotopy associativity of {H-spaces} {I.,II.}
\newblock {\em Trans. Amer. Math. Soc.}, {\bf 108}(1963), 275--312.

\bibitem{GS}
M. Gerstenhaber and S. D. Schack. \newblock Bialgebra Cohomology,
Deformations, and Quantum Groups.
\newblock {\em Proc. Nat. Acad.
Sci.}, {\bf 87}(1990), 478--481.

\end{thebibliography}
\end{document}